\documentclass[10pt]{article}

\usepackage[margin=1.0in]{geometry}
\usepackage{authblk}
\usepackage{amsmath}
\usepackage{graphicx}
\usepackage{parskip}
\usepackage{xcolor}
\usepackage{hyperref}

\newcommand\mplnck{M_\mathrm{Planck}}

\title{Boson stars in $f(T)$ extended theory of gravity}
\author{S.~Iliji{\'c}}
\author{M.~Sossich}
\affil{University of Zagreb,
       Faculty of Electrical Engineering and Computing,\\
       Department of Applied Physics, Unska 3, HR-10\,000 Zagreb, Croatia}

\date{\today}

\begin{document}

\maketitle

\begin{abstract}
Spherically symmetric configurations
of the non-interacting massive complex scalar field,
representing non-rotating boson stars,
are considered within the framework of the
modified torsion based $f(T)$ gravity,
with $f(T) = T + \alpha \, T^2/2$.
We find that with sufficiently large negative value of $\alpha$
the mass of the boson stars can be made arbitrarily large.
This is in contrast to General Relativity where an upper bound,
$M_{\max} \sim \mplnck^2/m$, to the mass of the boson stars
built from the non-interacting scalar field exists
and where the masses of boson stars in the astrophysical regime
can be obtained only with the introduction
of the scalar field self-interaction.
With sufficiently large negative $\alpha$
we also find negative gravitational binding energy for all masses,
which can be seen as an indication of the stability of such configurations.
In its positive regime, $\alpha$ can not be made arbitrarily large
as a phase transition in the stress--energy components of the
$f(T)$-fluid develops.
This phenomenon has already been reported to occur
in polytropic stars constructed within the $f(T)$ gravity theory.
\end{abstract}


\section{Introduction \label{sec:intro}}

Recent cosmological observations \cite{Riess_1998, Perlmutter_1999},
as well as the ever open quest
for the complete quantum theory of gravity \cite{kieferqg, schulz2014review},
motivate the research in the field of modified theories of gravity.
One of the common ways to modify General Relativity (GR)
is to allow for a nonlinear function $f$ of the scalar curvature $R$
in the Einstein--Hilbert action.
The resulting modified theories of gravity based on curvature
are known as $f(R)$ gravity theories \cite{Sotiriou:2008rp, NOJIRI201159, NOJIRI20171}.
They have been successfully applied
from cosmological settings \cite{PhysRevD.88.124036, PhysRevD.95.103519, capozzello2019}
to high curvature gravity regimes \cite{PhysRevD.16.953, HASSLACHER1981221, PhysRevD.51.4250}.
However, It is well known that a theory of gravity
equivalent to General Relativity (GR)
can be formulated in terms of torsion instead of curvature
\cite{Aldrovandi:2013wha}.
This theory replaces the scalar curvature
of the Einstein--Hilbert action by the torsion scalar $T$,
it uses the tetrad field as the dynamical degree of freedom,
and is known as the teleparallel equivalent of GR (TEGR).
In the same spirit as the $f(R)$ gravity theories
modify the curvature--based GR,
a nonlinear function $f$ can be used to modify the action of TEGR.
The resulting theories of gravity, known as $f(T)$ gravity theories,
have recently gained considerable attention.
While most of applications of the $f(T)$ gravity are in the field of cosmology
\cite{%
   deAndrade:2000kr,
   Yang:2010hw,
   Myrzakulov:2010vz,
   Cai:2015emx, 
   Iorio:2015rla, 
    Farrugia:2016xcw, 
   Awad:2017yod,
   Awad:2017ign,
   Bahamonde:2017ize,    
   Qi:2017xzl,
   Ferraro:2018tpu, 2019IJMPD..2830016C},
the applications dealing with the static spherical symmetry
are somewhat less in number \cite{Wang:2011xf,Daouda:2011rt, 2017EPJP..132..250G, Velay-Vitow:2017odc, 2017EPJC...77...62P, 2017EPJC...77..283P}.
An important problem found in the early formulations of $f(T)$ gravity
was the lack of Lorentz invariance
in the sense that the equations of motion were not invariant
with respect to the particular choice of the tetrad fields,
regardless of the latter satisfying
the expected metric compatibility condition.
This problem was pointed out and investigated by many authors
\cite{%
Blagojevic:2000pi,
   Li:2010cg,
   Sotiriou:2010mv,
   Li:2011rn,
   Bejarano:2014bca, 
    Krssak:2015oua,  
   Nester:2017wau,  
   Hohmann:2018rwf},
and is still not fully understood \cite{Golovnev:2017dox, universe5060158}.
In this work we will rely on the covariant formulation of $f(T)$ gravity
as proposed by Kr{\v s}{\v s}{\'a}k and Saridakis \cite{Krssak:2015oua}.
As the particular form of $f$ we will use $f(T) = T + \alpha T^2 / 2$,
since this form of $f$ guarantees the correct GR-limit
when the parameter $\alpha \to 0$.


Static spherically symmetric vacuum solutions in $f(T)$ gravity
have been considered in \cite{Xin-H-2011, Ahmed:2016cuy, DeBenedictis:2016aze}
,
while the solutions involving the polytropic fluid
and Yang--Mills field have been considered in \cite{PhysRevD.98.064047}
and \cite{PhysRevD.98.064056}.
In this work we will construct static spherically symmetric
self-gravitating configurations of the non-interacting complex scalar field.
Our motivation to study this matter model
comes in one part from its relative simplicity,
while in the other part it comes from the fact
that scalar field is the key component
of the standard $\Lambda$CDM model of cosmology.
It is therefore interesting to explore objects
that could be constructed of self gravitating scalar fields
in the primordial or in any other cosmological epoch.
The self-gravitating configurations of the scalar field
are in GR commonly referred to as boson stars \cite{Schunck08,Liebling12}.
The maximal mass of a Boson star
formed by the non-interacting scalar field
is in GR estimated to be $M_{\max} \sim \mplnck^{2}/m$,
where $m$ is the scalar field mass,
the estimate being based on the assumption
that the scalar field is confined within a radius
comparable to the Compton wavelength
and that it is bound by the uncertainty principle
and gravity \cite{Ruffini69,Liebling12}.
Such configurations are sometimes called mini--Boson stars \cite{JETZER1992163}.
Boson stars that have masses that are in the astrophysical regime
can in GR be obtained only with the introduction of
the scalar field self--interaction \cite{PhysRevLett.57.2485}.

In this work we will numerically construct Boson stars
in $f(T) = T + \alpha T^2 / 2$ gravity.
We will first inspect the global parameters of the solutions,
such as the gravitational mass and the particle number,
and compare these to the corresponding values in GR.
We will also inspect the radial profiles
of the energy density and the principal pressures,
as these may reveal features that are specific to the  $f(T)$ gravity theory.
The paper is organized as follows:
in Sec.~\ref{sec:foft} we introduce the $f(T)$ gravity action
and derive the field equations specific
to the scalar field in the static spherical symmetry.
In Sec.~\ref{sec:bcond} we discuss
the boundary conditions and our numerical procedure.
In Sec.~\ref{sec:masspart} we compute
the gravitational mass and the particle number
for boson stars in $f(T)$ gravity
with positive and negative values of the parameter $\alpha$
and we compare these to the GR-case.
In Sec.~\ref{sec:phasetrans} we investigate
a specific feature that develops in the radial profiles
of the energy density and the principal pressures
as $\alpha$ reaches a critical value in its positive regime.
We conclude the paper in Sec.~\ref{sec:concl}.
We use natural units $c=1=\hbar$ throughout the paper so that $G=1/\mplnck^2$,
and we use the metric signature $\eta_{\mu\nu}=\mathrm{diag}(1,-1,-1,-1)$.


\section{Field equations in static spherical symmetry in $f(T)$ \label{sec:foft}}

The $f(T)$ gravity theory action can be written as
\begin{equation} \label{eq:action}
   S = \int \left(
       \frac{f(T)}{16\pi G} + \mathcal{L}_{\mathrm{matter}}
   \right) h \; \mathrm{d}^{4}x ,
\end{equation}
where $f$ is in general a nonlinear function of the torsion scalar $T$,
$h$ is the determinant of the tetrad $h^a{}_\mu$,
and $\mathcal{L}_{\mathrm{matter}}$
is the Lagrangian density due to matter fields.
We use Latin symbols for the tetrad indices,
and Greek symbols for the spacetime indices.
The tetrad satisfies the metric compatibility condition
$h^a{}_\mu h^b{}_\nu g^{\mu\nu} = \eta^{ab}$,
where $g^{\mu\nu}$ is the spacetime metric tensor
and $\eta^{ab}$ is the metric of Minkowski.
If $f(T)=T$, the variation of the action (\ref{eq:action})
with respect to the tetrad gives field equations
that are equivalent to those of general relativity (GR)
and the resulting theory of gravity is known
as the teleparallel equivalent of GR (TEGR).
If $f$ is nonlinear in $T$,
the field equation resulting from the variation of (\ref{eq:action})
with respect to the tetrad can be written as
\begin{equation} \label{eq:genericeom}
h^{-1} h^{a}{}_{\mu} \partial_{\sigma}
     \left( h \frac{\mathrm{d} f(T)}{\mathrm{d} T} S_{a}{}^{\nu\sigma} \right)
- \frac{\mathrm{d} f(T)}{\mathrm{d} T} T_{\alpha\beta\mu} S^{\alpha\beta\nu}
+ \frac12 f(T) \delta_{\mu}{}^{\nu}
+ \frac{\mathrm{d} f(T)}{\mathrm{d} T}
          S_{a}{}^{\alpha\nu}
          h^{b}{}_{\mu}
          \omega^{a}{}_{b\alpha}
 = 8\pi G \mathcal{T}_{\mu}{}^{\nu},
\end{equation}
where
\begin{equation}\label{eq:stress}
   \mathcal{T}_{a}{}^{\mu} = - \frac{1}{h}
         \frac{\delta(h\mathcal{L}_{\mathrm{matter}})}{\delta h^{a}{}_{\mu}}
   = - \frac{\partial \mathcal{L}_{\mathrm{matter}}}{\partial h^{a}{}_{\mu}}
   - h_{a}{}^{\mu}\mathcal{L_{\mathrm{matter}}},
\end{equation} 
is the standard stress--energy tensor and
\begin{equation} \label{eq:torsiontensor}
T^{\alpha}{}_{\beta\gamma} =
h_{a}{}^{\alpha} \left( \partial_{\beta}  h^{a}{}_{\gamma}
                      - \partial_{\gamma} h^{a}{}_{\beta} \right)
+ h_{a}{}^{\alpha} \omega^{a}{}_{b\beta}  h^{b}{}_{\gamma}
- h_{a}{}^{\alpha} \omega^{a}{}_{b\gamma} h^{b}{}_{\beta}
\end{equation}
is the torsion tensor.
The quantity $\omega^{a}{}_{b\alpha}$
is the inertial spin connection which is,
in the covariant formulation of $f(T)$ gravity
proposed by Kr{\v s}{\v s}\'ak and Saridakis \cite{Krssak:2015oua},
determined from the requirement that the torsion tensor vanishes
in the flat space limit of the metric.
%
%
%
The tensors
\begin{equation} \label{eq:ktensor}
    K_{\alpha\beta\gamma} = \frac12 \left(  T_{\alpha\gamma\beta}
      + T_{\beta\alpha\gamma} +  T_{\gamma\alpha\beta} \right)
\end{equation}
and
\begin{equation} \label{eq:stensor}
    S_{\alpha\beta\gamma} =  K_{\beta\gamma\alpha}
     + g_{\alpha\beta} \,  T_{\sigma\gamma}{}^{\sigma}
      - g_{\alpha\gamma} \,  T_{\sigma\beta}{}^{\sigma}
\end{equation}
are known as the contorsion tensor
and the modified torsion tensor respectively.
The torsion scalar,
\begin{equation}
T =T^{\alpha\beta\gamma} S_{\alpha\beta\gamma},
\end{equation}
is then defined as the contraction of the torsion tensor
with the modified torsion tensor.

In order to study the most simple form of $f(T)$ we chose to work with
\begin{equation} \label{eq:alphafoft}
f(T)=T+\frac \alpha 2 T^{2},
\end{equation}
where the parameter $\alpha$ is allowed
to have both positive and negative values.
In the limit $\alpha \to 0$
the resulting $f(T)$ gravity theory reduces to TEGR,
implying the equivalence of the resulting field equations with those of GR.
For nonzero $\alpha$, terms proportional to $\alpha$
appear in the field equation (\ref{eq:genericeom}),
and for sufficiently small values of $\alpha$,
the solutions to the field equations are expected
to differ continuously from their GR counterparts.
The departure of the solutions from the well known GR solutions
is expected to reveal features of $f(T)$ gravity theory
which we aim to study in this paper.
The form (\ref{eq:alphafoft}) can also be seen
as the lowest two terms in the power expansion
of a more general nonlinear $f(T)$ having the correct GR-limit as $\alpha\to0$.

In order to facilitate the comparison of the solutions
obtained within $f(T)$ gravity to those of GR,
it is convenient to write the field equations
in the form that allows for the ``GR-picture interpretation''.
The field equations can be written as
\begin{equation} \label{eq:eomeff}
 G_{\mu}{}^{\nu}
 = 8\pi G \, \mathcal{T}_{\mathrm{eff}} {}_{\mu}{}^{\nu}
 = 8\pi G \, (\mathcal{T}_{\mu}{}^{\nu} + \mathcal{\tilde{T}}_{\mu}{}^{\nu}),
\end{equation}
where $G_{\mu}{}^{\nu}$ on the l.h.s.\ is the Einstein's tensor of GR,
while on the r.h.s.~we introduce the effective stress--energy tensor
$\mathcal{T}_{\mathrm{eff}} {}_{\mu}{}^{\nu}$
as the sum of the the matter stress--energy tensor (\ref{eq:stress})
and $\mathcal{\tilde{T}}_{\mu}{}^{\nu}$ (denoted with the tilde)
which consists of terms proportional to $\alpha$.
The tensor $\mathcal{\tilde{T}}_{\mu}{}^{\nu}$
can therefore be interpreted as the stress--energy introduced
by the nonlinearity of $f(T)$ or the stress--energy of the ``$f(T)$-fluid''.

As we intend to work in static spherical symmetry,
we use spherical coordinates $x^{\mu} = (t,r,\vartheta,\varphi)$
and write the tetrad field as
\begin{equation} \label{eq:tetrad}
  h^{a}{}_{\mu} = \mathrm{diag} \left(
  \mathrm{e}^{\Phi(r)}, \mathrm{e}^{\Lambda(r)}, r, r \sin \theta
  \right),
\end{equation}
which through the metric compatibility condition
implies the static spherically symmetric metric
\begin{equation} \label{eq:metric}
  g_{\mu\nu} = h^{a}{}_{\mu} h^{b}{}_{\nu} \, \eta_{ab}
  = \mathrm{diag} \left( \mathrm{e}^{2\Phi(r)},
     - \mathrm{e}^{2\Lambda(r)}, - r^2, - r^2 \sin^2\vartheta \right).
\end{equation}
The flat space limit of the above metric is obtained
by letting the metric profile functions $\Phi(r)\to0$ and $\Lambda(r)\to0$.
The condition that the components of the torsion tensor
vanish in the flat space limit
allows one to construct the spin connection $\omega^a{}_{b\alpha}$.
Its nonzero components are found to be
$\omega^{\hat r}{}_{\hat \vartheta \vartheta}
= - \omega^{\hat \vartheta}{}_{\hat r \vartheta} = -1$,
$\omega^{\hat r}{}_{\hat \varphi \varphi}
= - \omega^{\hat \varphi}{}_{\hat r \varphi} = - \sin\vartheta$, and
$\omega^{\hat \vartheta}{}_{\hat \varphi \varphi}
= - \omega^{\hat \varphi}{}_{\hat \vartheta \varphi} = -\cos\vartheta$
(coordinate labels are used as indices
and the orthonormal ones are denoted with the hat symbol).
The resulting torsion scalar is
\begin{equation}
  T = \frac{ 2 \mathrm{e}^{-2 \Lambda (r)}
      \left( \mathrm{e}^{\Lambda (r)}-1 \right)
      \left(\mathrm{e}^{\Lambda (r)}-2 r \Phi '(r)-1\right) }{ r^2 } ,
\end{equation}
where prime denotes differentiation with respect to $r$.

The matter Lagrangian for the non-interacting (free)
complex scalar field $\phi$ is
\begin{equation}
 \mathcal{L}_{\mathrm{matter}}
 = \frac12 g^{\mu \nu} \left(
     \nabla_{\mu}\phi^* \nabla_{\nu}\phi
   + \nabla_{\mu}\phi \nabla_{\nu}\phi^* \right) - m^2 \phi^* \phi,
\end{equation}
where $\nabla_{\mu}$ is the covariant derivative and $m$ is the field mass.
According to (\ref{eq:stress}),
the stress--energy tensor of the scalar field is
\begin{equation}
  \mathcal{T}_{\nu}{}^{\mu}
    = \nabla_{\nu}\phi^{*} \nabla^{\mu}\phi
    + \nabla_{\nu}\phi \nabla^{\mu}\phi^{*}
    -\delta_{\nu}^{\mu} \left(
       \nabla^{\sigma}\phi^{*} \nabla_{\sigma}\phi - m^2 \phi^*\phi
    \right).
\end{equation}

For the scalar field we use the standard time-stationary harmonic ansatz
compatible with the assumed static spherical symmetry,
\begin{equation}
   \phi(r,t) = \phi(r) \, \mathrm{e}^{ - \mathrm{i} \omega t },
\end{equation}
where from here on $\phi(r)$ denotes a real profile function
depending on the radial coordinate only.
The constant $\omega$ can be interpreted as the energy
of the quantum of the scalar field.
The above field ansatz avoids the instability problems
given by the Derric's Theorem \cite{Derrick:1964ww},
while for the other methods see e.g.\ \cite{Diez13, TDLee87}.

In order to write down the field equations of $f(T)$ gravity theory
in the GR-picture (\ref{eq:eomeff}),
we start with with the well known components of the Einstein tensor,
\begin{alignat}{1}
G_t{}^t & = r^{-2} \left( 1 - \mathrm{e}^{-2\Lambda}
            \left( 1 - 2 r \Lambda' \right) \right) , \label{eq:eintt} \\
G_r{}^r & = r^{-2} \left( 1 - \mathrm{e}^{-2\Lambda}
            \left( 1 + 2 r \Phi' \right) \right) , \label{eq:einrr} \\
G_\vartheta{}^\vartheta & = G_\varphi{}^\varphi
            = r^{-1} \mathrm{e}^{-2\Lambda} \left(
            \left( \Lambda' - \Phi' \right)
            \left( 1 + r \Phi' \right) - r \Phi'' \right) \label{eq:einang} ,
\end{alignat}
and proceed to the components of the stress--energy tensor
$\tilde{\mathcal{T}}_{\mu}{}^{\nu}$ of the $f(T)$--fluid which can be given by
\begin{alignat}{1}
  8\pi G \tilde{\mathcal{T}}_{t}{}^{t} & = - \alpha r^{-4}
   \mathrm{e}^{-4 \Lambda } \left(\mathrm{e}^{\Lambda }-1\right)
   \Big(\left(\mathrm{e}^{\Lambda }-1\right)
   \left(\left(\mathrm{e}^{\Lambda }-5\right) \left(\mathrm{e}^{\Lambda
   }-1\right)-4 r^2 \left(2 \Phi ''+\Phi
   '^2\right)\right) \notag \\
   & \qquad + 4 r \Lambda ' \left(3
   \left(\mathrm{e}^{\Lambda }-1\right)+2 \left(\mathrm{e}^{\Lambda
   }-3\right) r \Phi '\right)\Big)
   \label{eq:setfofttt} \\
  8\pi G  \tilde{\mathcal{T}}_{r}{}^{r} & = - \alpha r^{-4}
   \mathrm{e}^{-4 \Lambda } \left(\mathrm{e}^{\Lambda }-1\right)
   \left(\mathrm{e}^{\Lambda }-2 r \Phi '-1\right)
   \left(\left(\mathrm{e}^{\Lambda }-1\right) \left(\mathrm{e}^{\Lambda
   }+3\right)+2 \left(\mathrm{e}^{\Lambda }-3\right) r \Phi
   '\right) 
   \label{eq:setfoftrr} \\
  8\pi G \tilde{\mathcal{T}}_{\vartheta}{}^{\vartheta} & =
  8\pi G \tilde{\mathcal{T}}_{\varphi}{}^{\varphi} = \alpha r^{-4}
   \mathrm{e}^{-4 \Lambda } \Big(\left(\mathrm{e}^{\Lambda }-1\right)
   \Big(\left(\mathrm{e}^{\Lambda }+3\right) \left(\mathrm{e}^{\Lambda
   }-1\right)^2 \notag \\
   & \qquad + 2 r \Big( \Phi' \left(\mathrm{e}^{\Lambda }-2
   \mathrm{e}^{2 \Lambda }-4 r^2 \Phi ''+1\right) -2 r^2 \Phi
   '^3+3 \left(\mathrm{e}^{\Lambda }-1\right) r \Phi ''+3
   \left(\mathrm{e}^{\Lambda }-1\right) r \Phi
   '^2\Big)\Big) \notag \\ & \qquad + 2 r \Lambda ' \left(r \Phi '
   \left(2 \left(2 \mathrm{e}^{\Lambda }-3\right) r \Phi '-3
   \left(\mathrm{e}^{\Lambda }-3\right) \left(\mathrm{e}^{\Lambda
   }-1\right)\right)-3 \left(\mathrm{e}^{\Lambda
   }-1\right)^2\right)\Big)
   \label{eq:setfoftang}
\end{alignat}
Finally, the stress--energy components due to the scalar field are given by
\begin{alignat}{1}
 \mathcal{T}_{t}{}^{t} & = \mathrm{e}^{-2 \Lambda } {\phi'}^2+\phi ^2
   \left(m^2+\mathrm{e}^{-2 \Phi } \omega ^2\right) = \rho ,
   \label{eq:setphitt} \\
 \mathcal{T}_{r}{}^{r} & =
   - \mathrm{e}^{-2 \Lambda } {\phi'}^2
   + \phi^2 \left(m^2-\mathrm{e}^{-2 \Phi } \omega ^2\right)
   = - p , \label{eq:setphirr} \\
 \mathcal{T}_{\vartheta}{}^{\vartheta} &
   = \mathcal{T}_{\varphi}{}^{\varphi} = \mathrm{e}^{-2 \Lambda } {\phi'}^2
   + \phi ^2 \left(m^2-\mathrm{e}^{-2 \Phi } \omega ^2\right) = - q,
   \label{eq:setphiang}
\end{alignat}
and can be interpreted in terms of the energy density $\rho$,
the radial pressure $p$ and the transverse pressure $q$ of the boson fluid.

The variation of the action (\ref{eq:action})
with respect to the scalar filed
gives the field equation $\nabla^{\mu}\nabla_{\mu}\phi-m^2\phi=0$,
which can be written out as
\begin{equation} \label{eq:phieqn}
    r \phi \left( \omega ^2 \mathrm{e}^{-2 \Phi}-m^2 \right)
  + \mathrm{e}^{ -2 \Lambda }
    \left( \phi' \left(-r \Lambda' + r \Phi'+2\right) + r \phi''\right ) = 0.
\end{equation}
The above differential equation, which has the same structure as in GR,
together with the three independent differential equations
that follow from (\ref{eq:eomeff})
and whose parts are given by (\ref{eq:eintt})--(\ref{eq:setphiang}),
complete the set of equations to be satisfied
by the tetrad profile functions $\Phi(r)$ and $\Lambda(r)$
and the scalar field profile function $\phi(r)$.


\section{Boundary conditions and the numerical procedure \label{sec:bcond}}

In order for the solutions
to the field equations derived in the preceding section
to represent non-rotating boson stars,
they must be global and must satisfy certain boundary conditions.
As $r \to \infty$, one expects the energy density
and the pressures of the boson fluid to vanish
and the spacetime metric to approach
that of the static spherically symmetric vacuum
of the gravity theory that is being considered.
As can be seen from (\ref{eq:setphitt})--(\ref{eq:setphiang}),
vanishing of the stress--energy components of the boson fluid
implies $\phi \to 0$ and $\phi' \to 0$.
The static spherically symmetric vacuum 
in $f(T)$ gravity with $f$ given by (\ref{eq:alphafoft})
is asymptotically flat \cite{DeBenedictis:2016aze}
so the boundary conditions on the metric profile functions
are $\Phi \to 0$ and $\Lambda \to 0$.
At $r = 0$, in GR one requires
orthonormal components of the Riemann tensor to be finite,
which leads to the boundary conditions
$\Lambda(0) = 0$, $\Phi'(0 )= 0$, and $\Lambda'(0) = 0$.
These conditions apply also in the case of $f(T)$ gravity
with $f$ given by (\ref{eq:alphafoft}),
which can be shown by expanding the field equations (\ref{eq:eomeff})
in powers of $r$ near $r=0$.
One finds the conditions
\begin{alignat}{1}
   \alpha \, \mathrm{e}^{-4 \Lambda (0)}
     \big(\mathrm{e}^{\Lambda(0)}-1\big)^3
     \big(\mathrm{e}^{\Lambda(0)}-5\big) \, {r^{-4}}
   + \mathcal{O} \left({r^{-3}} \right) & = 0, \\
   \alpha \, \mathrm{e}^{-4 \Lambda (0)}
     \big(\mathrm{e}^{\Lambda (0)}-1\big)^3
     \big(\mathrm{e}^{\Lambda (0)}+3\big) \, {r^{-4}}
   + \mathcal{O} \left({r^{-3}} \right) & = 0,
\end{alignat}
where the leading terms are due to the nonlinearity of $f$.
The above conditions are satisfied at $r=0$ if $\Lambda(0)=0$,
which is exactly one of the condition known form GR.
Plugging $\Lambda(0)=0$ into higher-order terms of the same power expansion,
as well as into the power expansion of the field equation (\ref{eq:phieqn}),
one finds the conditions
\begin{alignat}{1}
  4 \Lambda'(0)
    \Big( 1 + 2 \alpha  \Lambda'(0) \big( \Lambda'(0) - 2 \Phi'(0) \big) \Big)
    \, r^{-1} + \mathcal{O} \big( r^0 \big) & = 0, \\
 \big( \Lambda '(0)-\Phi '(0)\big)
    \Big( 1 + 2 \alpha  \Lambda '(0) \big(\Lambda '(0)-2 \Phi '(0)\big) \Big)
    \, {r^{-1}} +\mathcal{O} \big( r^0 \big) & = 0, \\
 \phi '(0) \, {r^{-1}} + \mathcal{O} \big(r^0\big) & = 0,
\end{alignat}
that are satisfied if
$\Lambda'(0)=0$, $\Phi'(0)=0$ and $\phi'(0)=0$,
which are equivalent to the boundary conditions known from GR.
Interestingly, another solution to the above conditions exists
as a consequence of introducing
the $T^2$ term in (\ref{eq:alphafoft}) and which appears to be
$\Phi'(0)={\Lambda'(0)}/{2}+{1}/{(4 \alpha \Lambda'(0))}$.
However, we will consider only the first result
as it corresponds to the GR case.
The problem with the second case is that in the limit $\alpha\rightarrow 0$,
i.e.\ in the GR-limit, $\Phi$ diverges at $r\to0$.
This case could be interesting in the regime where $\alpha$ is large,
but this theory would considerably deviate from GR
and would not pass the standard Solar system tests.
Summarizing the derived boundary conditions,
at $r=0$ we have $\Lambda=0$ and
$\Phi' = \Lambda' = \phi' = 0$ for the derivatives
(regularity at the origin),
while as $r\to\infty$ we have $\Phi\to0$, $\Lambda\to0$,
and $\phi\to0$ (asymptotic flatness).

A quantity of interest in any asymptotically flat solution in GR
is its gravitational or ADM mass, $M$,
which can, in a static spherically symmetric spacetime,
be interpreted as the mass of the central body (star).
In the modified theory, using the GR-picture it can be expressed as
\begin{equation}
   M = 4 \pi \int_{0}^{\infty} r^2 \rho_{\mathrm{eff}} \, \mathrm{d} r,
     = 4 \pi \int_{0}^{\infty} r^2 (\tilde\rho + \rho) \, \mathrm{d} r,
\end{equation}
where the effective energy density $\rho_{\mathrm{eff}}$
is the sum of the energy density
$\tilde\rho = \mathcal{\tilde T}_t{}^t$ of the $f(T)$-fluid
given by (\ref{eq:setfofttt})
and the energy density $\rho$ of the boson fluid
given by (\ref{eq:setphitt}).
Another quantity relevant to boson stars
is the particle number, $N$, which is given by
\begin{equation}
  N = \int j^0 \, h \, \mathrm{d}^3 x
= 4 \pi \int_{0}^{\infty} j^{0} \,
              \mathrm{e}^{\Lambda+\Phi} r^2 \, \mathrm{d} r,
\end{equation}
where $h = \det[h^a{}_\mu] = \sqrt{-\det[g_{\mu\nu}]}$,
and $j^0$ is the component of the conserved Noether current
implied by the $U(1)$ symmetry, which is given by
\begin{equation} \label{struja}
  j^{\mu} = \mathrm{i} \,
            ( \phi \, \partial^{\mu} \phi^* - \phi^* \partial^{\mu} \phi )
          = 2 \mathrm{e}^{-2\Phi} \omega \phi^2 \delta^{\mu}_0 .
\end{equation}
The mass of the boson star and the particle number
are both expected to be finite.

Due to the complexity of the field equations
we proceed to construct the solutions numerically.
We introduce dimensionless radial coordinate and variables,
\begin{equation}
   \tilde{r} = m r, \qquad
   \Omega = \frac{\omega}{m}, \qquad
   \tilde{\alpha} = \alpha m^2, \qquad
   \sigma = \sqrt{4 \pi G} \phi = \sqrt{4 \pi } \phi / \mplnck,
\end{equation}
and also the rescaled radial coordinate
\begin{equation}
   x = \frac{ \tilde{r} }{ \tilde{r} + 1 },
\end{equation}
mapping $0 \le \tilde r < \infty$ onto $0 \le x < 1$,
which is more appropriate for the numerical treatment.
Out of the four field equations presented in the preceding section
involving profile functions $\Phi$, $\Lambda$, and $\phi$ as unknowns,
only three are independent one of another.
We chose to work with the ${}^t_t$
and the ${}^\vartheta_\vartheta$-component of (\ref{eq:eomeff})
and with the field equation (\ref{eq:phieqn}),
as this choice allows simplest extraction
of the highest order derivatives $\Phi''$, $\Lambda'$, and $\sigma''$.
The ${}^r_r$-component of (\ref{eq:eomeff}) is used to verify the solutions.
The value $\sigma_0$ of the profile function $\sigma$ at $x=0$
and the value of the parameter $\tilde\alpha$
are used to parametrize the solutions,
while the a-priori unknown value of $\Omega$
(the rescaled scalar field frequency)
acquires the role of the eigenvalue of the boundary value problem.
Technically,
for the chosen values of $\sigma_0$ and $\tilde\alpha$
and a trial eigenvalue $\Omega$
we use power expansions of the rescaled field equations at $x=0$
to derive initial data at a point close to $x=0$
and evolve the equations in $x$ starting from that point.
Trial value of $\Omega$ is then fine-tuned
until boundary conditions as $x \to 1$ are satisfied.
This procedure allowed us to construct solutions
over a wide range of the parameter space.
Pairs of parameters $\sigma_0$ and $\tilde\alpha$ exist
for which we could obtain solutions having different eigenvalues $\Omega$
with zero or more nodes in $\sigma$.
Solutions with one or more nodes in $\sigma$
have been found in GR long ago \cite{Ruffini69}
and are usually referred to as excited boson stars.
They are generally considered to be unstable \cite{Lee:1988av},
so in this work we are considering only solutions with no nodes in $\phi$.


\section{Mass and particle number \label{sec:masspart}}

\begin{figure}
\begin{center}
\includegraphics{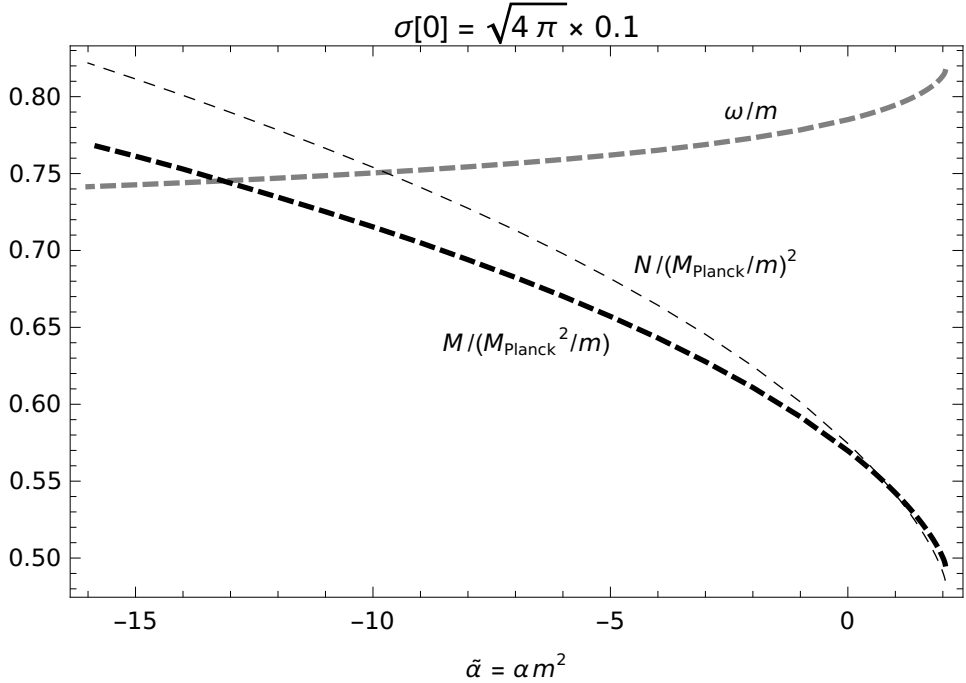}
\end{center}
\caption{\label{fig:one}
Gravitational mass $M$ (thick black dashed line,
in units of $\mplnck^2 m^{-1}$),
particle number $N$ (thin black dashed line, in units of $\mplnck^2 m^{-2}$),
and energy of scalar field quantum $\omega$
(thick gray dashed line, in units of $m$)
in boson stars with field mass $m$, no field self interaction,
central field amplitude $\sigma_0 = \sqrt{4\pi} \times 0.1$,
and a range of values of $\tilde\alpha = \alpha m^2$.}
\end{figure}

In order to obtain insight into the effects
that the $T^2$ term in (\ref{eq:alphafoft})
has on the structure of boson stars
we first generate a family of solutions that
has the fixed central value of the rescaled field profile function,
$\sigma(0) = \sigma_0 = \sqrt{4\pi} \times 0.1$,
while the value of $\tilde\alpha$ ranges over positive and negative values.
For each solution we obtain the eigenvalue $\Omega = \omega/m$
and compute the corresponding gravitational mass $M$
and the particle number $N$ of the boson star.
The results of this computation are shown in Fig.~\ref{fig:one}.
With $\tilde\alpha = 0$ we reproduce the well known GR solution.
With $\tilde\alpha$ going into the negative regime
the mass $M$ and the particle number $N$ are increasing,
while the eigenvalue $\Omega$ is decreasing.
We did not find any indication
that for even larger negative values of $\tilde\alpha$
than the ones shown in the figure the solutions would cease to exist
or that an upper bound on the mass would be reached.
With positive values of $\tilde\alpha$ less than a critical value,
$M$ and $N$ kept decreasing, while $\omega$ is increasing.
For the fixed value $\sigma_0 = \sqrt{4\pi} \times 0.1$ that we used,
the critical value of $\tilde\alpha$ is somewhat greater than $2$,
but since the stability of the numerical solutions becomes doubtful
as one approaches the critical $\tilde\alpha$,
its precise value could not be obtained.

The maximal mass of a stable boson star
built from non-interacting (free) scalar field in GR
is known to be small, $M_{\max} \sim \mplnck^{2}/m$,
and such boson stars are sometimes referred to as mini boson stars
\cite{JETZER1992163}.
However, according to another well known result \cite{PhysRevLett.57.2485},
the introduction of field self-interaction
allows for solutions with masses in the astrophysical range,
$M \sim \lambda^{1/2} M_{\mathrm{Ch}}=\lambda^{1/2}\mplnck^{3}/m^{2}_{H}$, where $m_{H}$ is the mass of the hydrogen atom.
Our result shows that with negative $\tilde\alpha$,
even without the field self-interaction,
the mass and particle number of a boson star
can become greater than that of a mini boson star in GR.

\begin{figure}
\begin{center}
\includegraphics{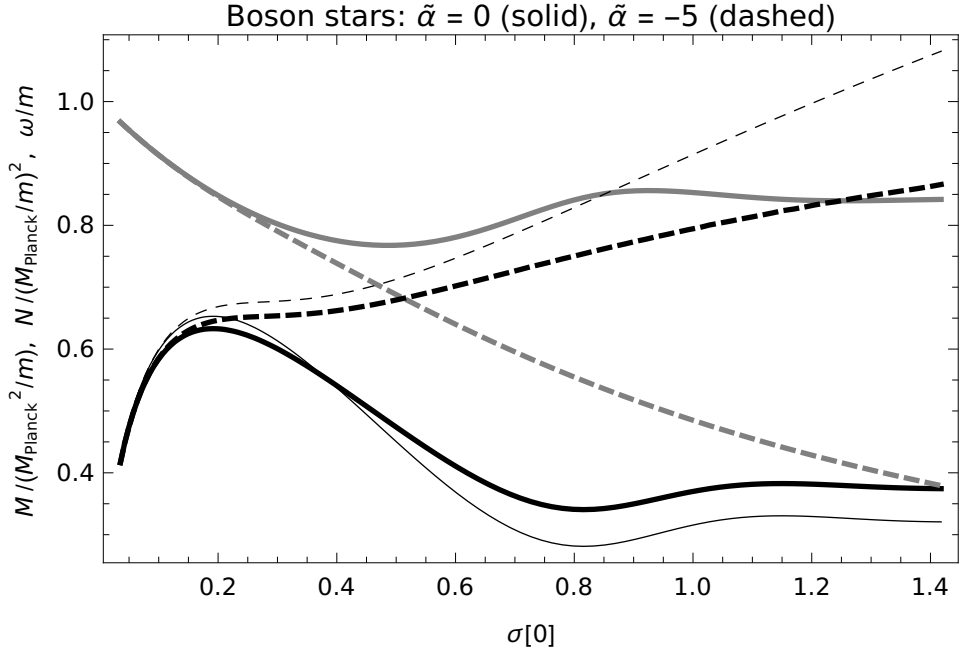}
\end{center}
\caption{\label{fig:two}
Gravitational mass $M$ (thick black lines),
particle number $N$ (thin black lines),
and energy of scalar field quantum $\omega$ (thick gray lines)
of boson stars with field mass $m$, no field self interaction,
a range of values of central field amplitude $\sigma_0 = \sqrt{4\pi} \phi_0$,
and $\tilde\alpha = 0$ (solid lines) and $\tilde\alpha = -5$ (dashed lines).}
\end{figure}

Our second family of solutions uses the fixed value $\tilde\alpha = - 5$
and varies the central value
of the rescaled field profile function $\sigma_0$.
Mass $M$, particle number $N$, and the eigenvalue $\Omega=\omega/m$,
are shown as functions of $\sigma_0$
in Fig.~\ref{fig:two} with dashed lines,
while the reference GR results ($\tilde\alpha = 0$) are shown with solid lines.
In GR, boson star mass $M$ and particle number $N$
increase with $\sigma_0$ up to the respective maxima,
followed by `damped oscillations'.
With $\tilde\alpha = -5$ no maximum in $M$ or $N$ was found
which can be considered as qualitatively different behavior.
This result is in line with our earlier finding that with constant $\sigma_0$
and increasing negative value of $\tilde\alpha$
no bound of $M$ or $N$ was found.

Another important feature one can observe in Fig.~\ref{fig:two}
is that in the case of GR, at a critical value of $\sigma_0$
somewhat greater than the one corresponding to the maxima of $M$ and $N$,
a crossing of the $M$-curve and the $N$-curve takes place.
This implies that at the critical configuration we have $M = Nm$,
$m$ being the mass parameter of the scalar field.
At values of $\sigma_0$ below the critical value, we have $M < N m$.
If the gravitational mass is understood as a sum of
the mass of the particles and the gravitational binding energy,
this would imply that gravitational energy is negative
which could further be understood as an indication
of the stability of the configurations.
At values of $\sigma_0$ greater than critical we have $M > N m$,
which would imply positive gravitational energy,
i.e.\ that positive work was done in order to bring
the infinitely dispersed particles into the given configuration.
Such configurations can hardly be imagined as stable.
Interestingly, in case of $\tilde\alpha = - 5$ we have $M < N m$
for all values of $\tilde\alpha$ that we have tested,
which can be seen as an indication of stability.
We provide further discussion of the issue of stability
of static spherically symmetric solutions in $f(T)$ gravity
in Sec.~\ref{sec:concl}.

\begin{figure}
\begin{center}
\includegraphics{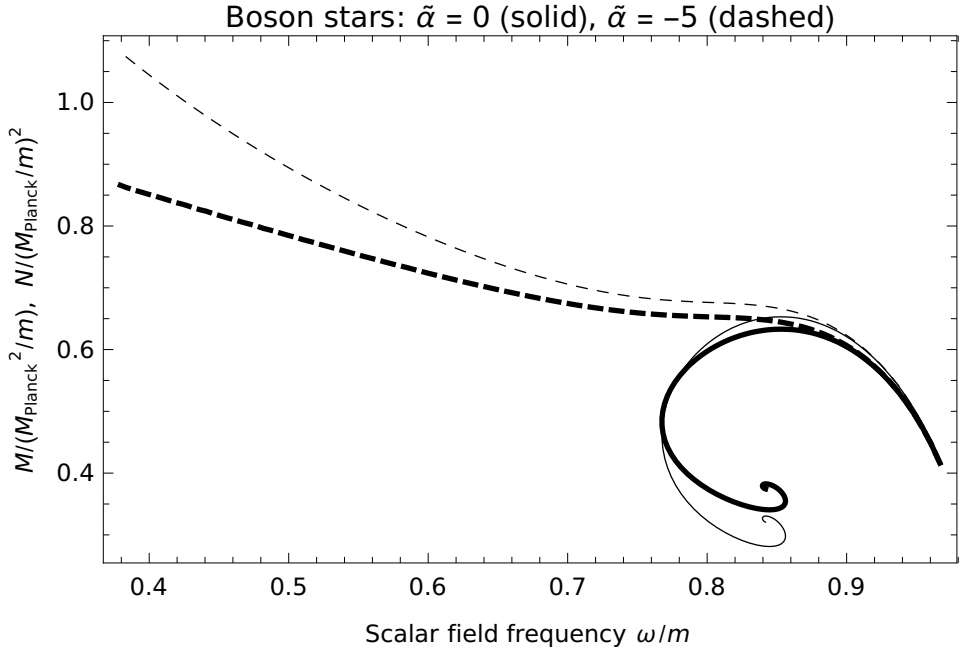}
\caption{\label{fig:three}
Gravitational mass $M$ (thick lines) and
particle number $N$ (thin lines) shown as functions of
energy of scalar field quantum $\omega$
in boson stars with field mass $m$, no field self interaction,
and $\tilde\alpha = 0$ (solid lines) and $\tilde\alpha = -5$ (dashed lines).}
\end{center}
\end{figure}

Fig \ref{fig:three} shows the dependence of the mass $M$
and the particle number $N$ of boson stars vs.\ the eigenvalue $\Omega$,
which corresponds to the rescaled energy quanta of the scalar field.
In GR case the solutions are always bounded within some range of $\omega$
where the minimal $\omega_{min}$ is present (solid lines).
With $\tilde\alpha = - 5$ the energy quanta are becoming arbitrary small,
even though the stellar mass is increasing.
In the case of positive $\tilde\alpha$ this feature is not observed
and the theory behaves like GR with the increased gravitational constant.

\section{Energy density and pressure profiles \label{sec:phasetrans}}

\begin{figure}
\begin{center}
\includegraphics{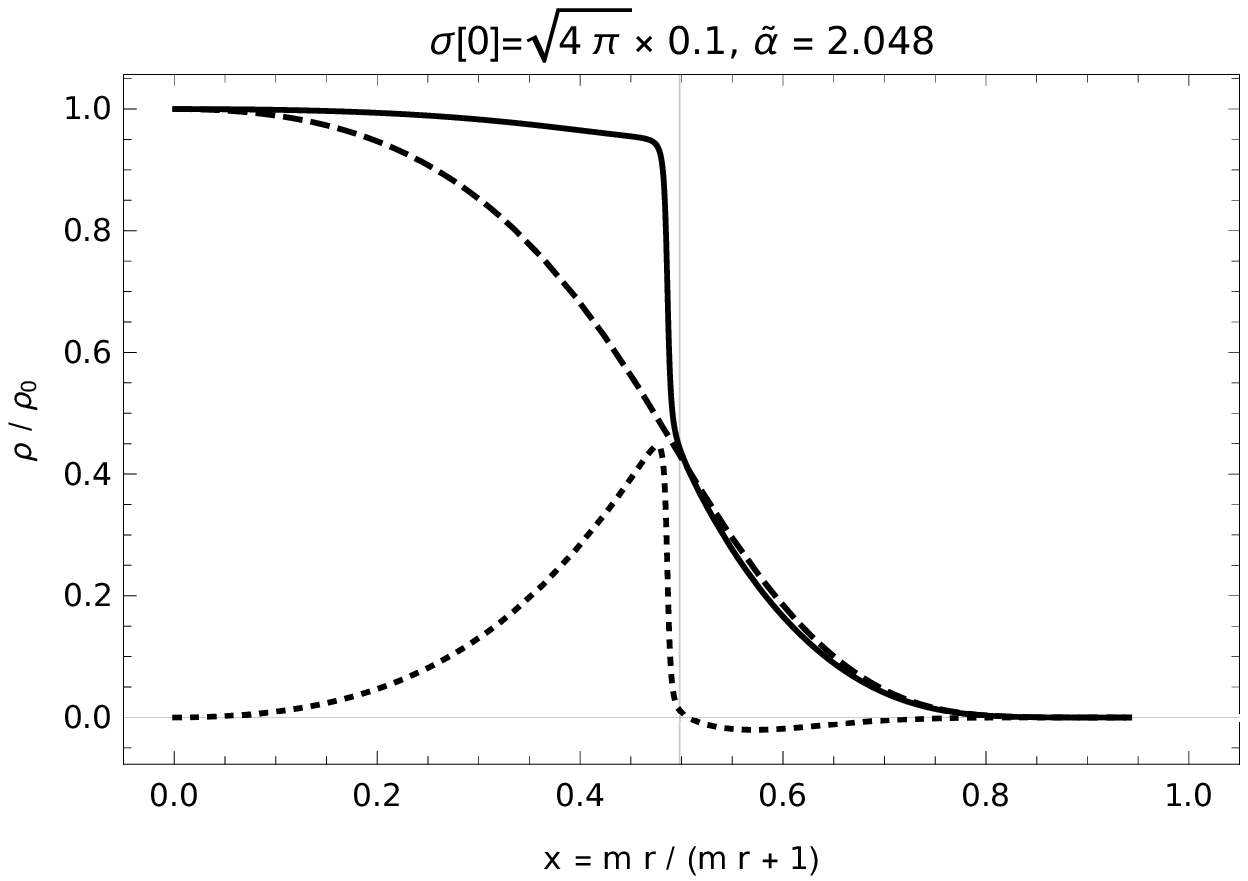}\\
\includegraphics{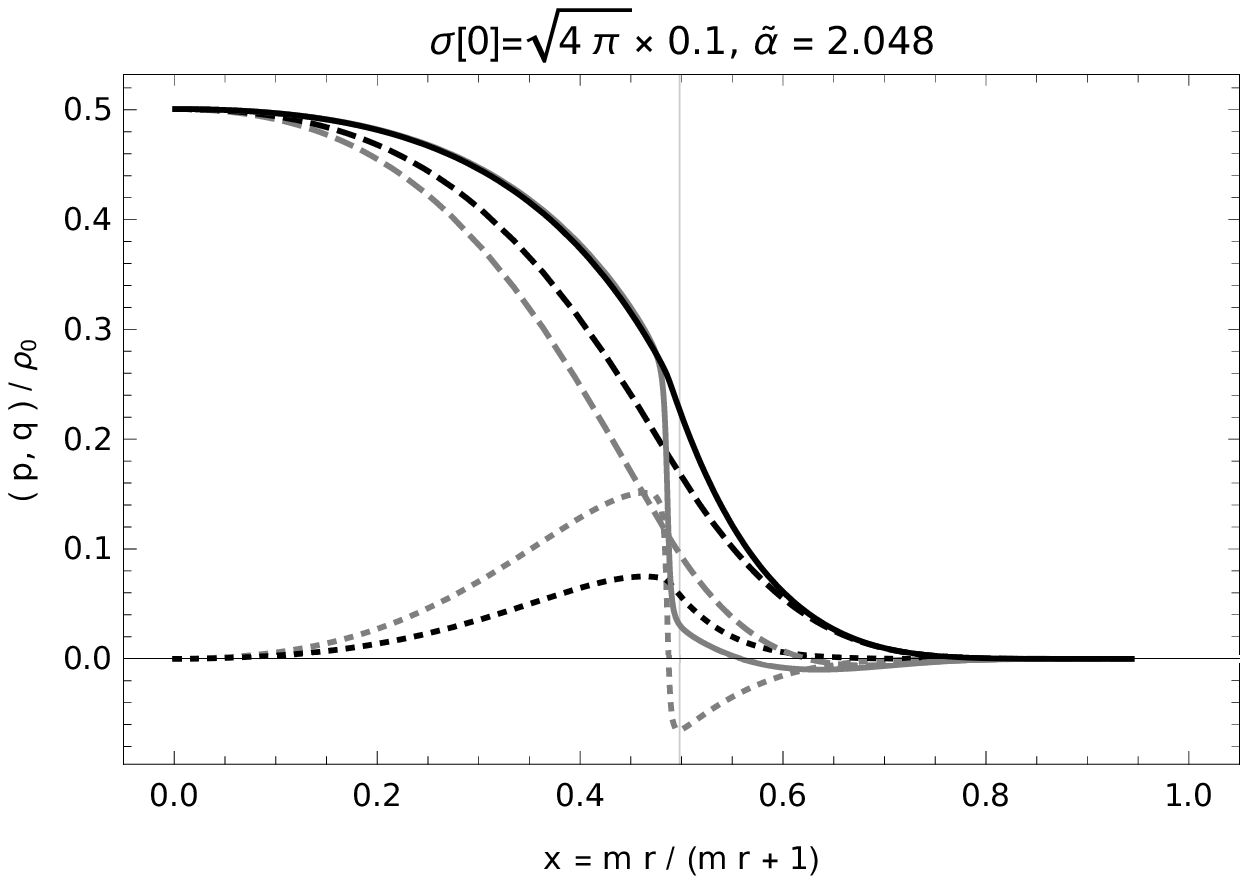}
\end{center}
\caption{\label{fig:four}
Radial profiles of stress--energy tensor components
for $\sigma_0 = \sqrt{4\pi} \times 0.1$ and $\tilde\alpha = 2.048$
(close-to-critical configuration).
Upper plot: effective energy density (solid line),
energy density due to boson fluid (long dashed line),
and energy density due $f(T)$-fluid (short dashed line).
Lower plot: effective radial and transverse pressures
(black and gray solid lines, respectively),
radial and transverse pressures due to boson fluid
(black and gray long dashed lines),
and radial and transverse pressures due to nonlinear terms in $f(T)$
(black and gray short dashed lines).
All quantities are shown relative to
value of effective energy density at $r=0$.
}
\end{figure}

The first family of solutions whose mass $M$, particle number $N$,
and the eigenvalue $\Omega=\omega/m$, are shown in Fig.~\ref{fig:one},
revealed that in the positive regime of the parameter $\tilde\alpha$
there is a critical value beyond which the solutions could not be obtained.
As one approaches the critical value of $\tilde\alpha$,
$M$ and $N$ approach zero, $\Omega$ approaches unity from below
(i.e.\ $\omega$ is approaching the scalar field mass parameter $m$),
and the solutions are becoming increasingly more difficult
to construct numerically.
In order to trace down the cause of this phenomenon
we looked into the radial profiles of the components
of the stress--energy tensor.
In the upper plot of Fig.~\ref{fig:four}
the radial profile of the effective energy density,
as well as the separate contributions
due to the boson fluid and due to the $f(T)$-fluid,
are shown for a solution with a close-to-critical value of $\tilde\alpha$.
The energy density of the boson fluid has a smooth outwardly decreasing profile
that is similar to the behavior this quantity in the GR solutions.
The energy density due to the $f(T)$-fluid is vanishing at the centre,
it is outwardly increasing up to a radially thin layer
within which it steeply drops to zero, becomes negative,
and asymptotically approaches zero (vacuum value) from the negative regime.
We will refer to this abrupt feature involving the change of sign
of the $f(T)$-fluid energy density as the ``phase transition''.
The effective energy density
(the sum of the energy densities of the boson and the $f(T)$-fluid)
is everywhere positive and outwardly decreasing,
but within the thin layer within which the phase transition
of the $f(T)$-fluid takes place it has a an abrupt step.

Lower plot of Fig.~\ref{fig:four}
shows the radial and the transverse effective pressures
as well as the separate contributions to these quantities
due to the boson fluid and the $f(T)$-fluid.
The radial and the transverse pressure profiles
of the $f(T)$-fluid vanish at the centre
and are outwardly increasing up to the phase transition layer,
where the radial pressure profile smoothly becomes outwardly decreasing,
while the transverse pressure abruptly changes sign.
In the most part of the interior of the boson star
the effective pressure appears to be isotropic (up to numerical noise),
regardless of both fluids having manifestly anisotropic pressures.
Outside of the phase transition layer effective pressure is anisotropic,
as is the static spherically symmetric vacuum solution in $f(T)$ gravity
theory \cite{DeBenedictis:2016aze}.

Attempts to increase the value of $\tilde\alpha$
beyond the value used in Fig.~\ref{fig:four}
made the steep portions of the radial profiles
of the energy density and the transverse pressure
of the $f(T)$-fluid even steeper.
It is therefore reasonable to expect that
the critical value of $\tilde\alpha$
corresponds to a discontinuity in these profiles.
It is also obvious that such solutions can not be obtained
by the numerical procedure we use.
We should also note that qualitatively similar phase transition
in the $f(T)$-fluid was found in \cite{PhysRevD.98.064047},
where compact stars composed of the polytropic fluid were considered.
It is therefore likely that with positive close-to-critical
values of $\tilde\alpha$ the phase transition described above is
a genuine feature of static spherically symmetric solutions
in the $f(T)$ gravity theory with $f$ given by (\ref{eq:alphafoft}).

Another interesting observation is that
if $M$ is the mass of the close-to-critical
(in the sense of $\tilde\alpha$ being as large as technically possible)
configuration of the boson star in $f(T)$,
the approximate value of the radial coordinate
at which the phase transition takes place
coincides with the value at which in GR
the horizon of the the Schwarzschild black hole of mass $M$ would form.
However, at this point we have no analytical arguments supporting
that above assertion.

\section{Conclusion \label{sec:concl}}

Recently, many modified theories of gravity have arisen
from the theoretical and experimental evidence that GR may be incomplete. 
In the desire to provide the fundamental description
of the nature of spacetime, which is the goal of any theory of gravity,
new views have been opened, in particular the concept of torsion.
Thus, the modified theories based on torsion
deserve attention as their curvature based counterparts do.
However, a modified theory of gravity that is to be taken seriously
must explain the complete spectrum of physical phenomena,
ranging from cosmological singularities, CMB, inflation,
to high energy physics, gravitational waves, black holes,
compact objects, etc. 
In this work we have explored the static spherically symmetric
self-gravitating configurations of the simplest bosonic matter
--- the non-interacting complex scalar field ---
within the framework of $f(T) = T + \alpha T^2 / 2$
torsion based modified theory of gravity.
Within GR, such solutions are known as boson stars.
In particular, when the non-interacting scalar field is used,
the mass of the star is bounded from above by $M_{\max} \sim \mplnck^2/m$,
and such stars are referred to as the mini boson stars,
while only with the introduction of the scalar field self-interaction
higher stellar masses can be obtained.

We have derived the field equations and the boundary conditions
relevant for the boson stars in $f(T)$ gravity
and we have numerically constructed the solutions
over a wide range of values of the parameters $\alpha$
and the central field amplitude $\phi_0$.
We have found that if $\alpha$ is negative and sufficiently large,
the mass of the boson is increasing with the central field amplitude $\phi_0$,
i.e.\ that it is no longer bounded as in GR.
This implies that within $f(T)$ with negative $\alpha$
boson stars formed of non-interacting scalar field
are not necessarily mini boson stars,
but may acquire masses in the astrophysical range.
The finding that the mass is no longer bounded might also be relevant
for the discussion of the dynamical stability of boson stars in $f(T)$,
since in GR the solutions with the central field amplitude
greater than the value corresponding to the maximal mass
were in some works found to be unstable
\cite{PhysRevD.38.2376, 1985MNRAS.215..575K, Gleiser:1988ih, Lee:1988av}.
Another finding that might be relevant for the discussion of the stability
of boson stars in $f(T)$ gravity
is that with sufficiently large negative $\alpha$
the binding energy of the boson star,
defined as the difference between its gravitational mass $M$
and the particle number $N$ multiplied by the field mass $m$,
is negative for all values of $\phi_0$ we have tested.
This implies that (positive) work must be applied
in order to disperse the boson fluid into individual particles
with negligible gravitational interaction, which indicates stability.
In GR, the binding energy of boson stars is negative
only for $\phi_0$ less than a critical value,
which is interpreted as onset of dynamical instability
\cite{Gleiser:1988ih}.
An analysis that could provide conclusive answers
to the problem of stability of boson stars within $f(T)$ gravity
would require perturbation of the time-dependent field equations
around the static solutions we have computed.
However, this endeavour lies outside of the scope of the present work
and we leave it for a future project.
With $\alpha$ in the positive regime,
we have found that the solutions can be obtained
only up to a critical value of $\alpha$.
In an attempt to trace down the cause of this phenomenon
we have adopted the ``GR-picture'' where the the features of the
$f(T)$ gravity theory are viewed as the presence of the stress energy tensor
of a quantity that we refer to as the ``$f(T)$-fluid'',
which together with the stress energy tensor of the scalar field
constitutes effective stress energy on the r.h.s.~of the Einstein equation.
As we approached the critical positive value of $\alpha$,
we could observe the development of abrupt sign changes
in the energy density and the transverse pressure of the $f(T)$-fluid,
eventually becoming step-like.
This phenomenon that we referred to as the ``phase transition''
has already been found in the context of polytropic stars
within $f(T)$ gravity.

We can conclude that the structure of boson stars
formed of non-interacting complex scalar field minimally coupled
to $f(T) = T + \alpha T^2 / 2$ gravity
is vastly different from its GR counterpart.
We believe that further research that could
include perturbative analysis of spherically symmetric
time-dependent field equations (stellar pulsations)
or axially symmetric solutions (rotating boson stars)
could reveal further interesting features
specific to torsion based $f(T)$ gravity.


\bibliography{marko}

\end{document}